\documentclass[12pt]{article}
\usepackage{graphicx}
\usepackage{epsfig}

\newcommand\pubnumber{DPF2013-171}
\newcommand\prenumber{FERMILAB-CONF-13-436-E}
\newcommand\pubdate{\today}

\def\napoli{Fermilab\\
Batavia, Illinois, USA}
\def\support{\footnote{Work supported by the Office of Science, 
          U.S. Department of Energy}}

\def\Title#1{\begin{center} {\Large #1 } \end{center}}
\def\Author#1{\begin{center}{ \sc #1} \end{center}}
\def\Address#1{\begin{center}{ \it #1} \end{center}}

\newcommand\pubblock{\rightline{\begin{tabular}{l} \pubnumber\\
  \prenumber\\       \pubdate  \end{tabular}}}
\newenvironment{Abstract}{\begin{quotation}  }{\end{quotation}}
\newenvironment{Presented}{\begin{quotation} \begin{center} 
             PRESENTED AT\end{center}\bigskip 
      \begin{center}\begin{large}}{\end{large}\end{center} \end{quotation}}

\def\beq{\begin{equation}}
\def\eeq#1{\label{#1}\end{equation}}
\def\eeqn{\end{equation}}

\def\beqa{\begin{eqnarray}}
\def\eeqa#1{\label{#1}\end{eqnarray}}
\def\eeqan{\end{eqnarray}}

\let\bar=\overbar

\def\Dslash{\not{\hbox{\kern-4pt $D$}}}
\def\dslash{\not{\hbox{\kern-2pt $\del$}}}

\def\msb{{\bar{\ssstyle M \kern -1pt S}}}

\begin{document}
\begin{titlepage}
\pubblock

\vfill
\Title{Measurement of $B$ Baryon Properties at CDF}
\vfill
\Author{ Patrick T. Lukens\support \\
representing the CDF Collaboration}
\Address{\napoli}
\vfill
\begin{Abstract}
We report on mass and mean life measurements of several ground state
$b-$baryons, using 9.6 $\textrm{fb}^{-1}$ of data from $p\bar p$ collisions at
$\sqrt{s}=1.96$ TeV, and
recorded with the Collider Detector at Fermilab.
Data collected with triggers designed to collect events with 
$J/\psi \rightarrow \mu^+ \, \mu^-$ candidates
and events with hadrons displaced from the beamline are used to 
measure the masses and lifetimes of $\Lambda_b, \Xi_b^-, \Xi_b^0$, and
$\Omega_b^-$.
The first evidence for the process
$\Omega_b^- \rightarrow \Omega_c^0 \, \pi^-$ is also shown.
The results supersede our previous measurements of these quantities.

\end{Abstract}
\vfill
\begin{Presented}
DPF 2013\\
The Meeting of the American Physical Society\\
Division of Particles and Fields\\
Santa Cruz, California, August 13--17, 2013\\
\end{Presented}
\vfill
\end{titlepage}
\def\thefootnote{\fnsymbol{footnote}}
\setcounter{footnote}{0}

\section{Introduction}
The quark model has had great success in describing
the spectroscopy of hadrons.  
In particular, this has been the case for the $D$ and 
$B$ mesons, where all of the 
ground states have been observed \cite{PDG}.
The spectroscopy of $c$-baryons also agrees well with the quark model, 
and a  rich
spectrum of baryons containing $b$ quarks is predicted \cite{Jenkins}.
The accumulation of large data sets from the Tevatron and LHC
has  made possible the observation of most of the
$b$ baryon ground states containing a single heavy 
quark \cite{D0_Xi_b}-\cite{LHCb_Xib}
and several excited states 
\cite{CDF_Sigma_b,CMS_Xib0,LHCb_LambdabStar}.

In this paper, we report the 
measurements of mass and lifetime for several $b$ baryons.
These measurements are made in
$p\overline{p}$ collisions 
at a center of mass energy of 1.96 TeV using the
Collider Detector at Fermilab (CDF II), 
and correspond to an integrated luminosity of
9.6 fb$^{-1}$.  These results supersede previous work by this collaboration.

The strategy of the analysis presented here 
is to demonstrate the reconstruction
and property measurements of the $\Lambda_b$, 
$\Xi^{-,0}_b$ and $\Omega^-_b$ as 
natural extensions of measurements that are made on 
better known $b-$hadron states obtained in the same data.
All measurements made here are performed on 
$B \rightarrow J/\psi \, K$ systems, to provide a large sample
for comparison to the baryon measurements.  

\section{Reconstruction Methods\label{sect:Detector}}

The CDF II detector has been described in detail elsewhere 
\cite{CDF_detector}.  
The analysis presented here  is based on events recorded with two
different trigger algorithms.
The first is dedicated to the collection of a 
$J/\psi \rightarrow \mu^+ \mu^-$ sample.
The second data set used is triggered by a system designed to collect
particle candidates 
that decay with lifetimes characteristic of heavy flavor hadrons
by selecting events containing tracks that are displaced from the
beamline.

The analysis of the data 
obtained with the $J/\psi$ trigger
begins with a selection of well-measured
$J/\psi \rightarrow \mu^+ \mu^-$ candidates.
The trigger requirements are confirmed by selecting
 events that contain two
 oppositely  charged  muon candidates.
Both muon tracks are required to have associated position
measurements in at least three layers of the silicon vertex detector and
a two-track invariant  mass within 80 MeV/$c^2$ of the
world-average  $J/\psi$ mass \cite{PDG}. 
This range was chosen for consistency with our earlier $b-$hadron mass
measurements \cite{CDF_B_mass}.
This data sample provides approximately $6.5\times 10^7$  $J/\psi$
candidates, measured with an average mass resolution of $\sim20$ MeV/$c^2$.
 

The reconstruction of $K^0_s$, $K^{*}(892)^{0}$, and $\Lambda$
candidates uses all tracks with $p_T \, >$ 0.4 GeV/$c$,
that are not associated with muons in the $J/\psi$ 
reconstruction or trigger tracks in the hadronic trigger data.  
Pairs of oppositely charged tracks are combined to identify these
neutral decay candidates, and silicon detector information
is not used. 
Candidate selection for these neutral states is 
based upon the mass calculated for each track pair,
after the appropriate mass assignment for each track and
the flight distance 
for $K^0_s$ and $\Lambda$ candidates.


For events that contain a $\Lambda$ candidate,
the remaining tracks 
are assigned the pion or kaon mass, and
$\Lambda \, \pi^-$ or $\Lambda \, K^-$ combinations 
are identified that are consistent with the decay process
$\Xi^- \rightarrow \Lambda \, \pi^-$ or 
$\Omega^- \rightarrow \Lambda \, K^-$.  
Tracks with $p_T$ 
as low as 0.4 GeV/$c$ are used for $\Xi^-$ reconstruction.
However, event simulation
motivates a requirement of $p_T(K^-) > 1.0$ GeV/$c$
for the $K^-$ daughters from $\Omega^-$ decay.

The $\Xi^-$ and $\Omega^-$ candidates used in the hadronic trigger
data set have an additional
fit performed with the three tracks  that simultaneously 
constrains the full final state vertex and
the $\Lambda$ and $\Xi^-$ or $\Omega^-$  masses of the 
appropriate track combinations.
This fit provides the best possible estimate of the hyperon momenta and decay
positions.
The result of this fit is used to define a helix that
serves as the seed for an algorithm that associates silicon detector
hits with the $\Xi^-$ or $\Omega^-$ trajectory that is predicted by the fit.  
Candidates with track measurements in at least one layer of the 
silicon detector
have excellent impact distance resolution (average of 60 $\mu$m) for the
charged hyperon track. 

The charmed hyperons are used to reconstruct 
charmed, strange baryons through the processes
$\Xi_c^0 \rightarrow \Xi^- \, \pi^+$,
$\Xi_c^+ \rightarrow \Xi^- \, \pi^+ \, \pi^-$, and
$\Omega_c^0 \rightarrow \Omega^- \, \pi^+$.  
Charmed hyperon candidates are required to have a $\Xi^-$ or $\Omega^-$ 
measured in the silicon detector and at least one $\pi^+$ with 
$p_T > 2.0$ GeV/$c$ and $|d|>100 \, \mu$m.  A good fit of the
full set of final state tracks is required that contains $\Lambda$ and $\Xi^-$
or $\Omega^-$ mass constraints.  We also require $p_T > 4.0$ GeV/$c$ and
$ct > 100 \, \mu$m for the charmed hyperon candidates.


For all $B$ candidates with a $J/\psi$ 
in the final state 
we require $p_T(B) > 6.0$ GeV/$c$ and the $p_T(h) > 2.0 $ GeV/$c$
where $h$ is whatever hadron appears in the final state with the $J/\psi$.
These requirements reduce combinatorial background.  
We also require $ct > 100 \, \mu$m and $|d| < 100 \, \mu$m,
to reduce prompt and poorly reconstructed candidates.
Finally, all final states are required to satisfy a fit that constrains the
decay topology and the mass of the $\mu^+ \, \mu^-$ pair to the 
nominal mass of the $J/\psi$.  The hadron tracks do not make use of any silicon
detector information.  In this way, all decay time information is derived solely
from the muons, and the decay time resolution will be the same for all
$B$ hadrons in this data set.

The hadronic trigger data provides a sample of $B$ baryons through
the decay channels $H_b \rightarrow H_c \, \pi^-$, where $H_b$ is a $b$
baryon and $H_c$ is a $c$ baryon.  
The selection requires $c$ baryon candidates that satisfy 
appropriate mass
ranges and a good final state
fit that constrains the hyperon and
charmed hyperon masses.
The $\pi^-$ candidates are required to 
have an electric charge opposite to the $\Lambda$ baryon number, and
to be consistent with the trigger 
by having $p_T > 2.0$ GeV/$c$ and $|d| > 100 \, \mu$m.
A requirement of $ct > 100 \, \mu$m and $|d| < 100 \, \mu$m on the 
full final state
reduces prompt and poorly reconstructed candidates.
The background under the $\Xi_b$ states is also reduced by restricting the
sample based on the measured decay time of the $\Xi_c$ to the range
$-2\sigma_{ct} \, < \, ct(\Xi_c) \, < \, 3 c\tau_0(\Xi_c) + 2\sigma_{ct}$ 
where $\sigma_{t}$ is the calculated uncertainty on the decay time and
$\tau_0(\Xi_c)$ is the nominal mean life.

\section{Particle Properties \label{sect:Properties}}
The mass and lifetime of the $B$ hadrons will be measured by a fit
that bins the data in decay time($ct$), but
not in mass.  
The probability distributions used for the mass fits 
contain a Gaussian term for the signal and a polynomial to describe the
background for each time bin.  Mean life is 
calculated by virtue of the fact that the 
fractional occupancy in each time bin implies 
a particular proper flight and resolution
combination.
The lower limit of decay time bin $n$ is given by
$\lambda_n = 
\lambda_1 - \lambda_0 \ln \left( \frac{N_b - n + 1}{N_b} \right )$,
where $\lambda_0$ is the initial value of the mean life used in the fit,
and $N_b$ is the number of time bins.  The lower limit of the first bin,
$\lambda_1$, is chosen to be 100 $\mu$m to remove the majority of 
the prompt background.
This method populates the time bins with an equal number of $B$ candidates 
based on the initial lifetime estimate. 

\begin{figure}[hbt]
\psfig{figure=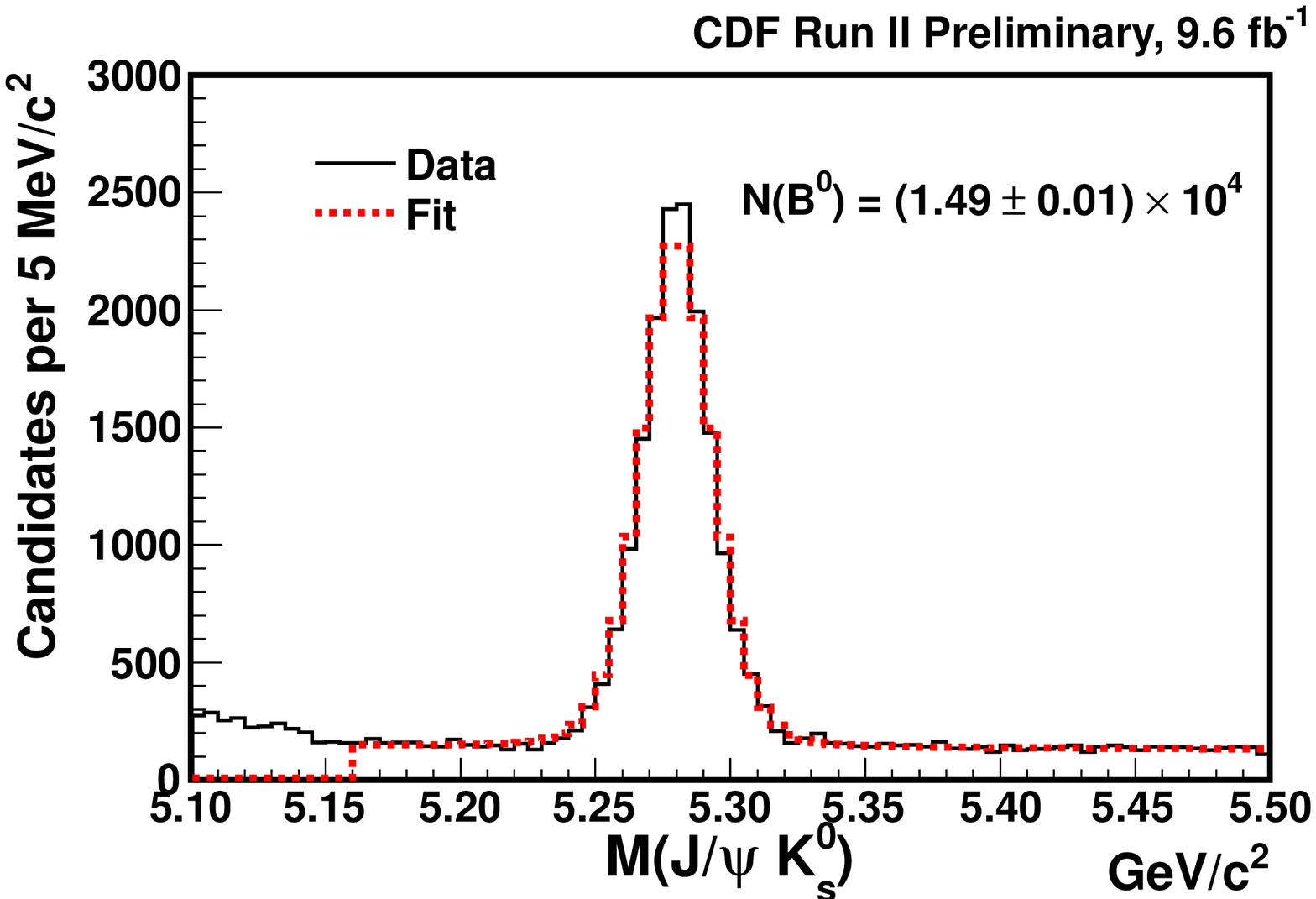,width=1.9in} 
\psfig{figure=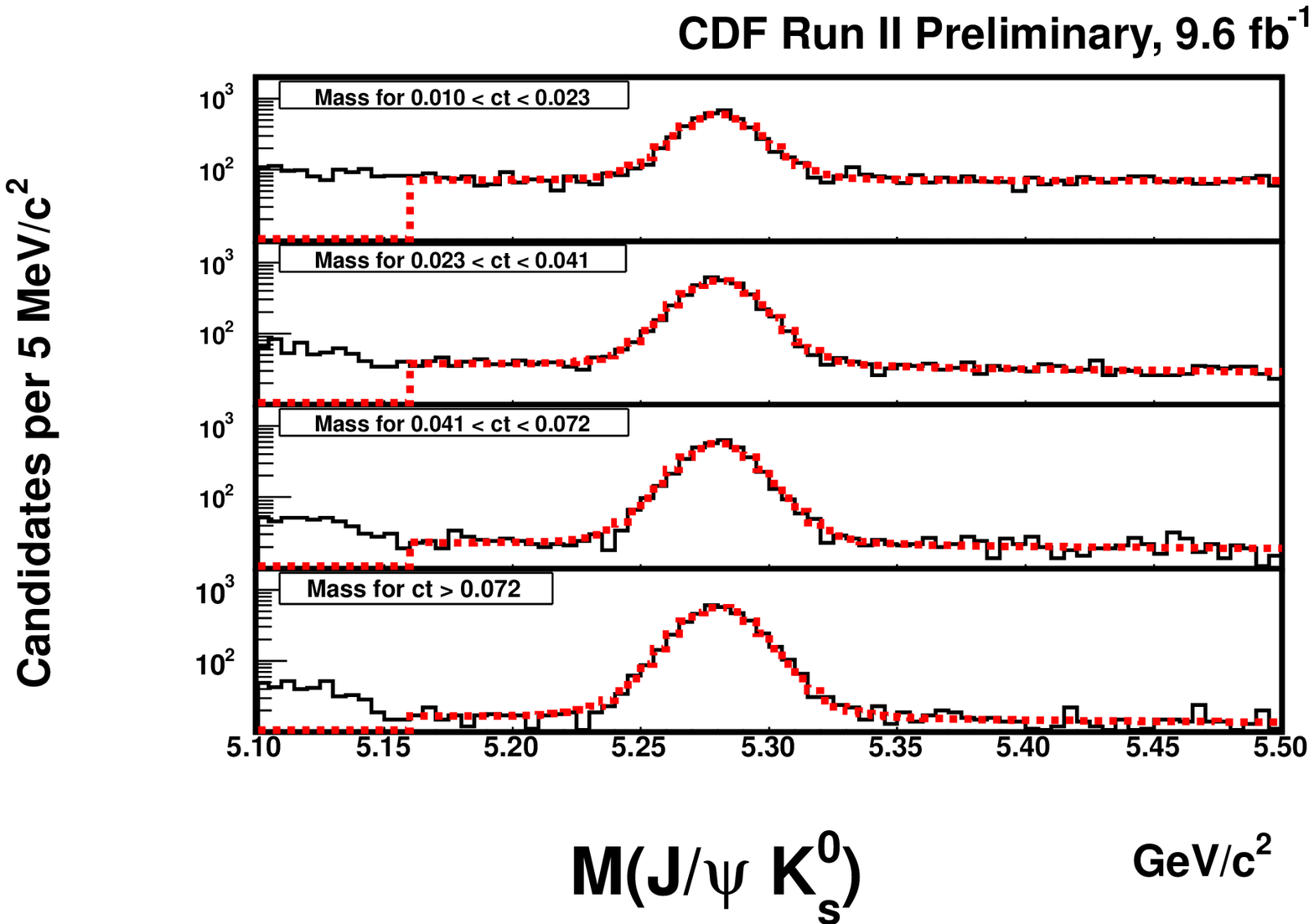,width=1.9in} 
\psfig{figure=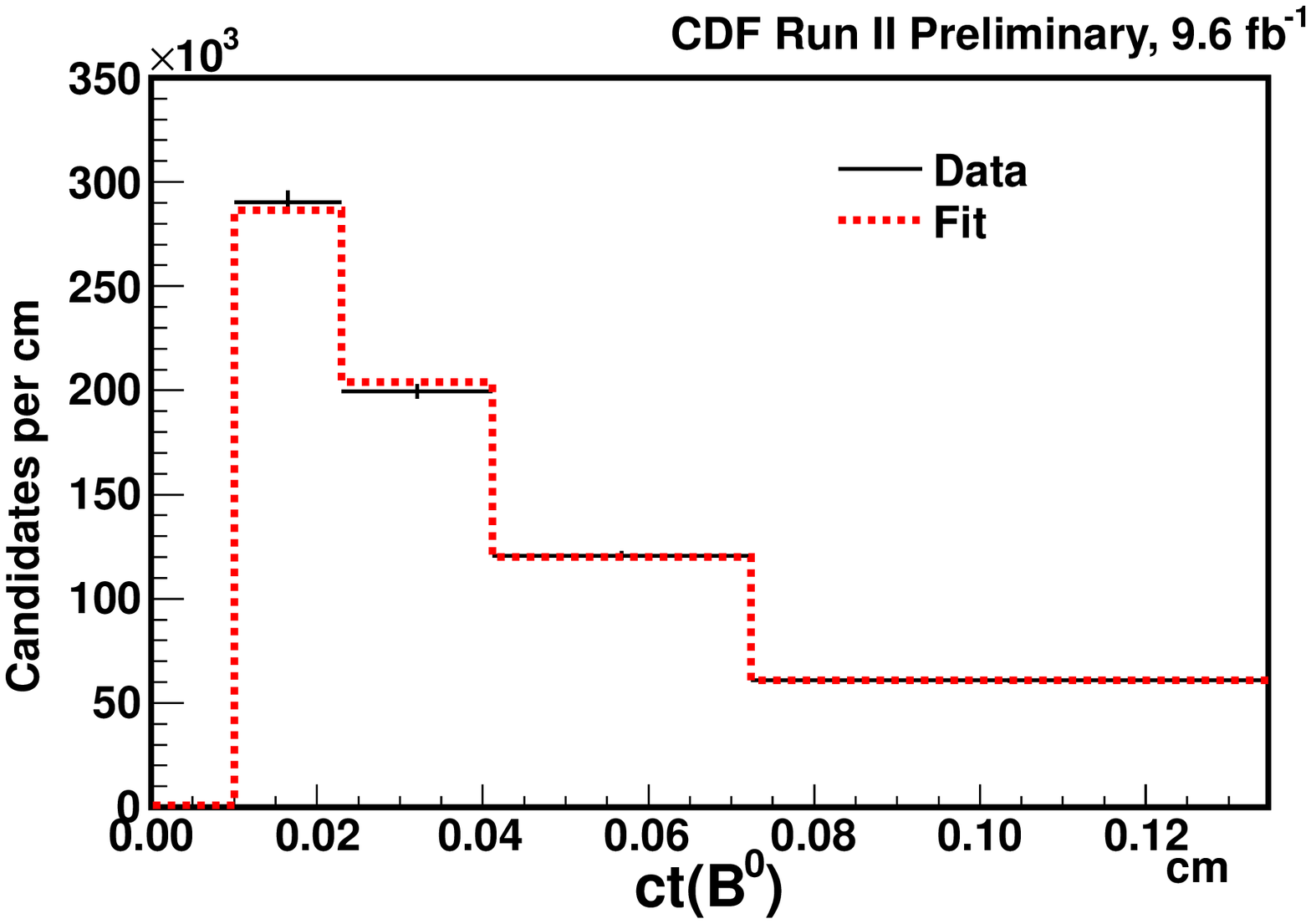,width=1.9in} 
\caption{The $J/\psi K^0_s$
mass distributions used for the strange $b$-meson mass 
and lifetime measurements.
The probability distributions obtained from the fits are 
overlaid on the data in dashed red.
 \label{fig:fig_17}}
\end{figure}
For the $B$ meson mass measurements, 
a single decay time bin is used. 
For the $B$ meson mean life measurements, 
$N_b = 4$ and $\lambda_0=$450 $\mu$m are used. 
An increase in the number of bins chosen
or a variation in the value of $\lambda_0$ by 10\%
is found to affect the measurement by less than 0.5\%.

The mass distributions and fit projections 
obtained for one $B$ meson reference signal is shown in Fig. \ref{fig:fig_17}.
The mass and mean life results obtained for the $B^+$ and $B^0$ are listed in 
Table \ref{table:compare_ref}.
Values are given for only the subset of the CDF Run II 
data that has not been included in the nominal value estimates \cite{PDG}.
These comparisons of the  unpublished data
with the nominal values serve as the calibration reference to
establish the systematic uncertainties on the mass and mean life measurements.

\begin{table}[hbt]
\begin{center}
\caption{$B$ Meson Mass and Mean Life Comparisons 
\protect \label{table:compare_ref}}
\vspace{3mm}
\begin{tabular}{cccccc}
\hline \hline
 Final State &  \multicolumn{2}{c}{Mass (MeV/$c^2$)}  & 
\multicolumn{2}{c}{Mean life ($\mu$m)} \\
\hline
\hline
 & Nominal & Difference & Nominal & Difference \\
  $J/\psi \, K^{+}   $  &  $5279.25\pm0.17$ & $0.7\pm0.2$ &  
$492.0\pm2.4$  & $-0.1\pm3.6$ \\
  $J/\psi \, K^{0*}  $  &  $5279.58\pm0.17$   & $0.6\pm0.2$ &  
$455.4\pm2.1$  & $3.6\pm5.1$  \\
 $J/\psi \, K^{0}_{s}$  &  $5279.58\pm0.17$  & $-0.5\pm0.2$ &  
$455.4\pm2.1$  & $1.2\pm6.1$ \\
\hline
\hline
\end{tabular}
\end{center}
\end{table}

\begin{figure}[hbt]
\psfig{figure=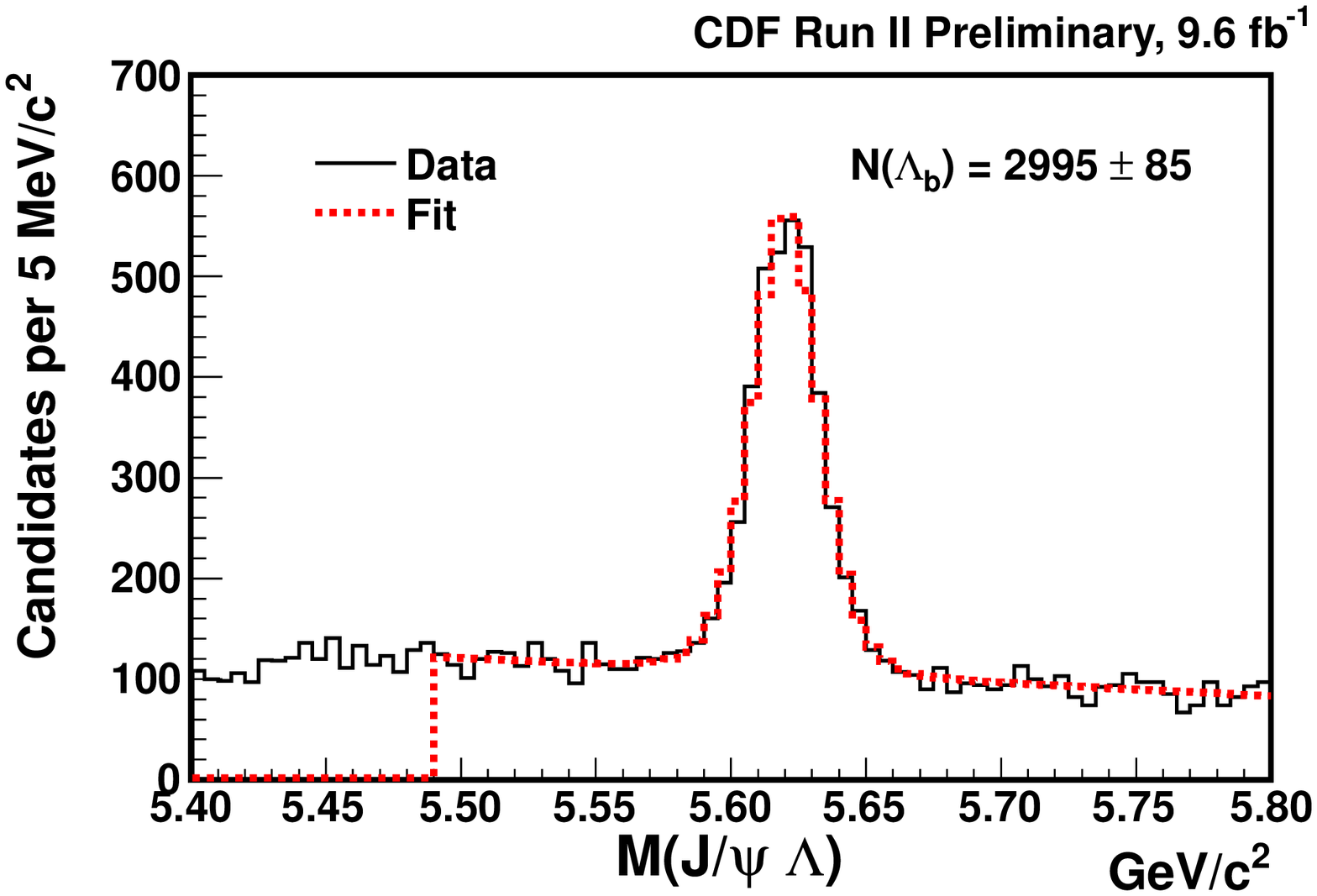,width=1.9in} 
\psfig{figure=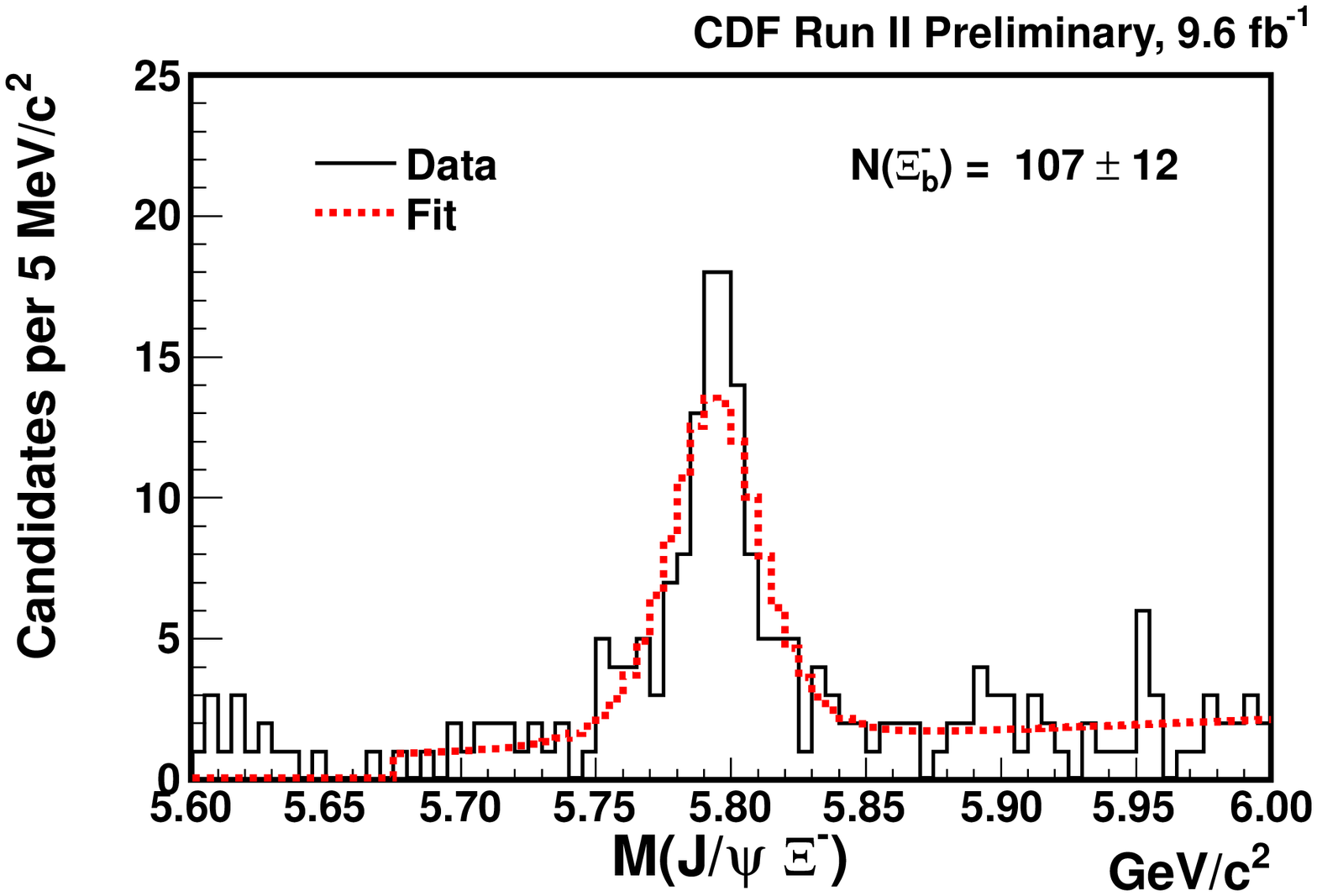,width=1.9in} 
\psfig{figure=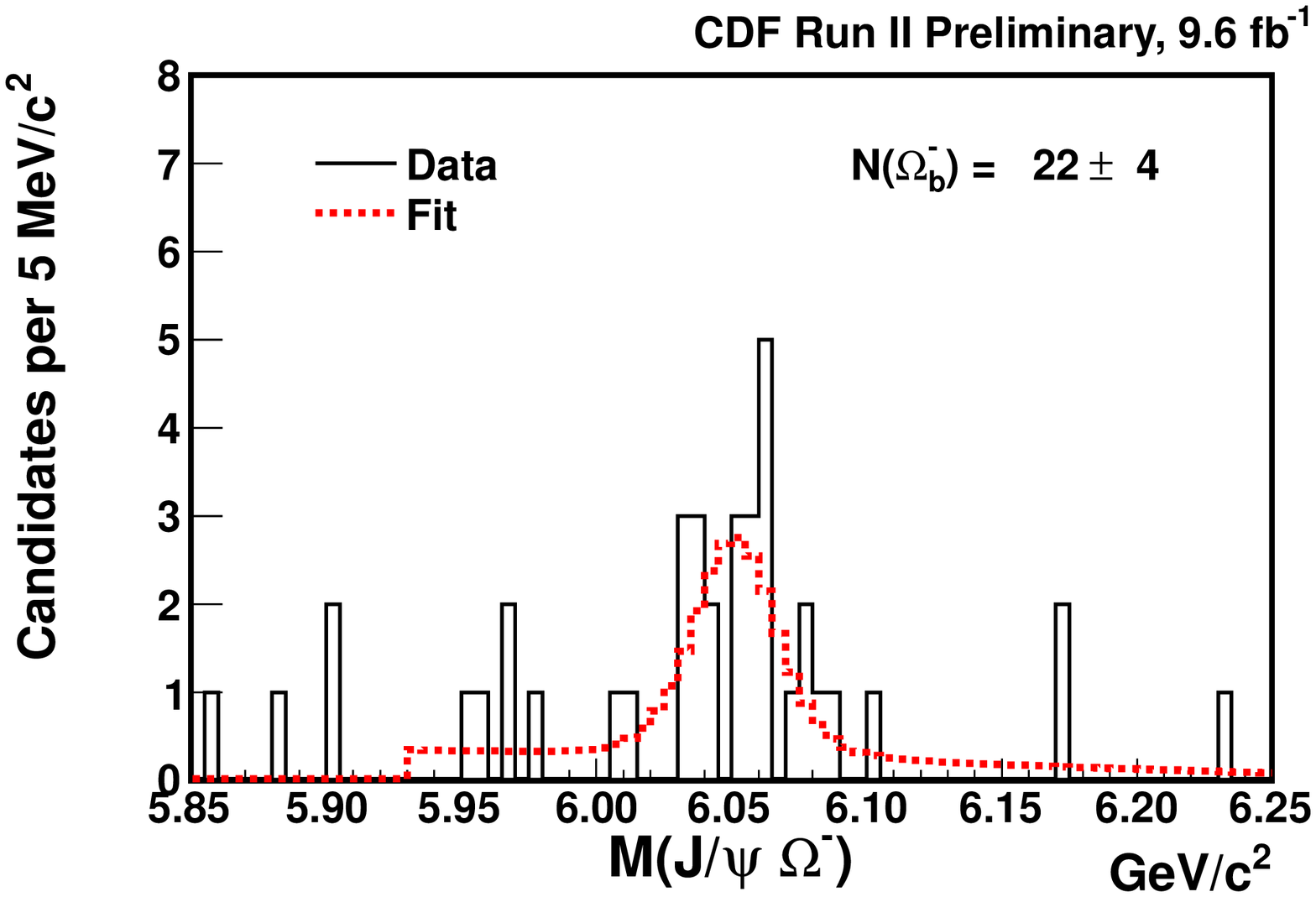,width=1.9in} 
\caption{The $J/\psi \Lambda$, $J/\psi \Xi^-$, and $J/\psi \Omega^-$
mass distributions used for the $b$-baryon mass measurements.
The probability distributions obtained from the fits are 
overlaid on the data in dashed red.
 \label{fig:fig_21}}
\end{figure}

The approach to fitting the mass and mean life of the $b-$baryons 
found in the $J/\psi$ sample is 
identical to that used for the meson reference signals.  
The mass distributions 
integrated in decay time and the 
projected fits are shown in Fig. \ref{fig:fig_21}.

\begin{figure}[hbt]
\psfig{figure=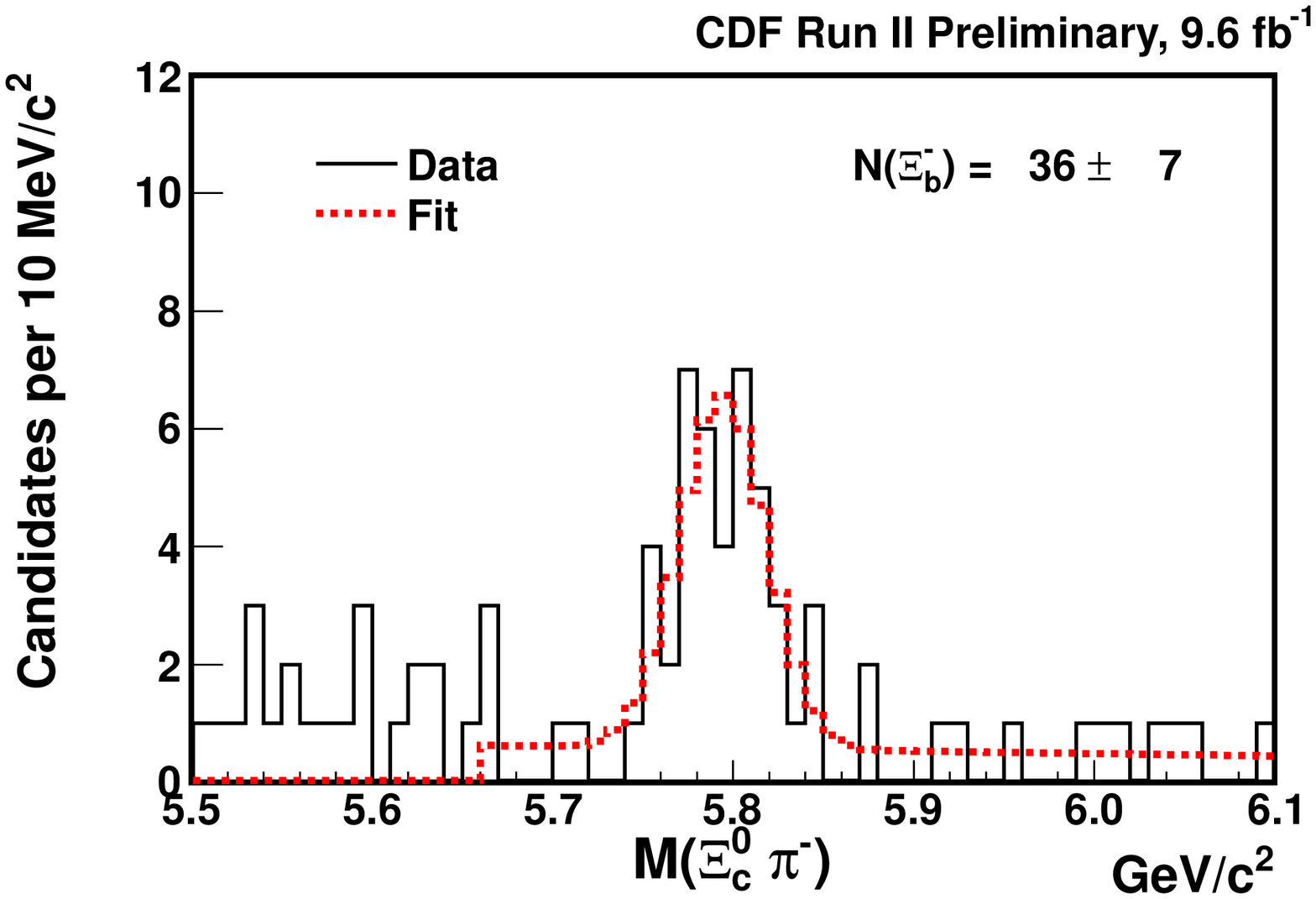,width=1.9in} 
\psfig{figure=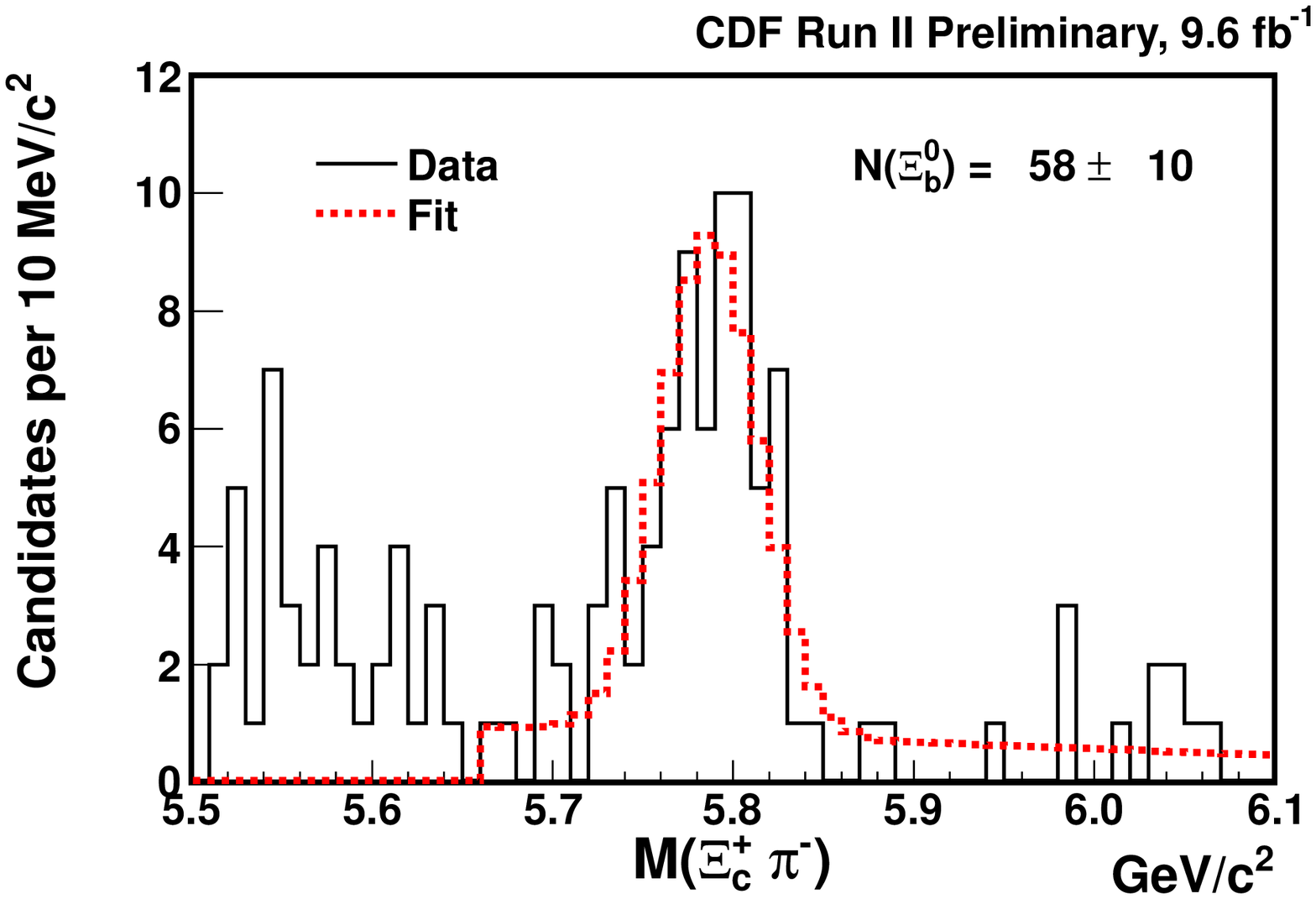,width=1.9in} 
\psfig{figure=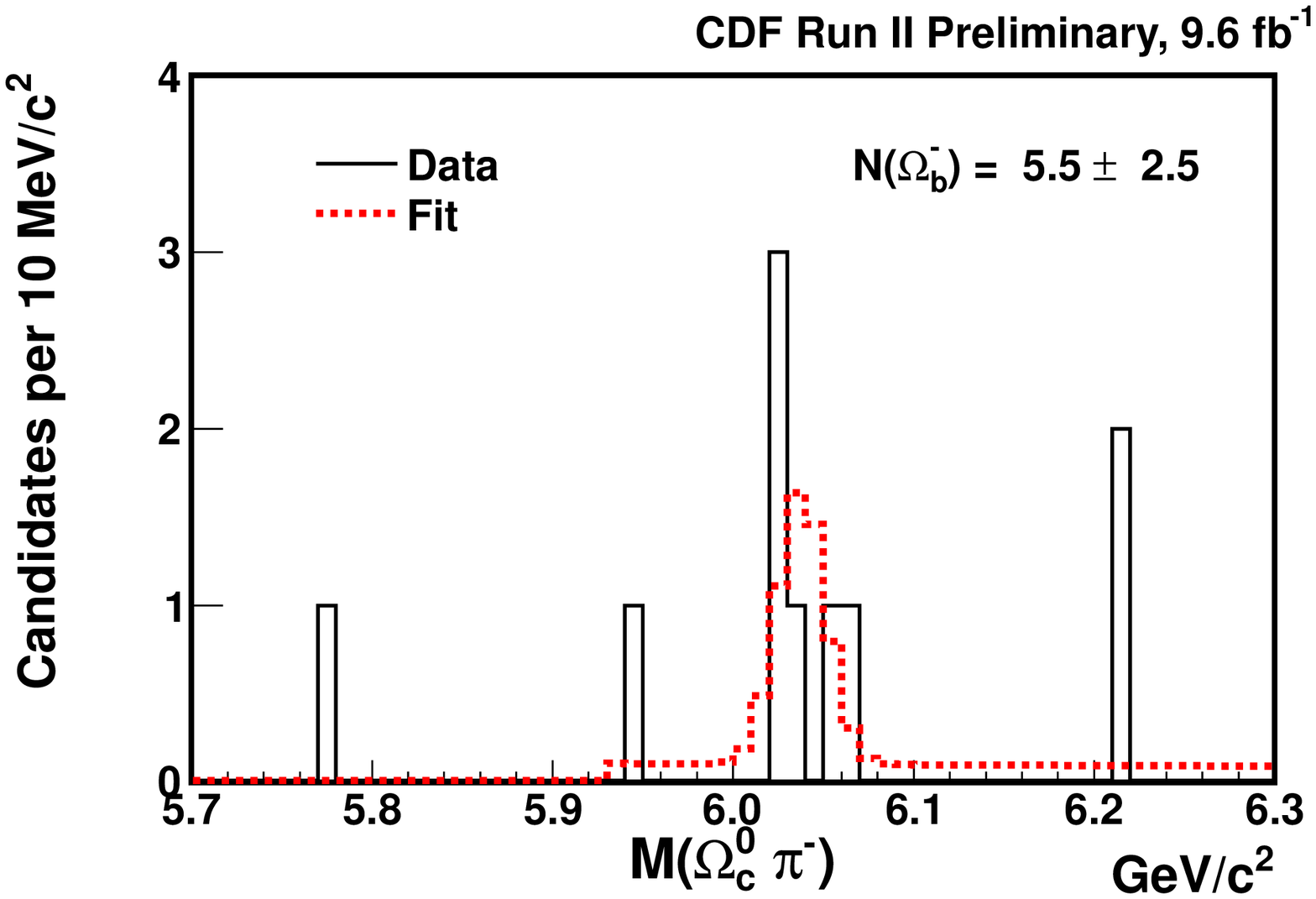,width=1.9in} 
\caption{The $\Xi_c^0 \pi^-$, $\Xi_c^+ \pi^-$, and $\Omega_c^0 \pi^-$
mass distributions used for the strange $b$-baryon mass measurements.
The probability distributions obtained from the fits are 
overlaid on the data in dashed red.
 \label{fig:fig_26}}
\end{figure}

The masses of the
$\Xi_b^-$ and $\Xi_b^0$ obtained from the $\Xi_b \rightarrow \Xi_c \, \pi^-$
processes are obtained by fitting the mass distributions.
These distributions and  projections of the fits
overlaid on the data are shown in Fig. \ref{fig:fig_26}.

There is a suggestion of an $\Omega_b^-$ signal in the $\Omega_c^0 \, \pi^-$
mass distribution
shown in Fig. \ref{fig:fig_26}.  Because this final state
has never been observed, the standard significance test was performed where
the fit was made with the null hypothesis and a floating amplitude.
The change in the fit likelihood, $\Delta 2 \ln {\cal L}$, was found to 
 correspond to 
a single sided fluctuation of a Gaussian distribution of $3.3 \sigma$.
While this does not meet the accepted standard for independent observation
of a signal, it indicates 
a very low probability of occurence due to a background fluctuation
and is evidence for
the process $\Omega_b^- \rightarrow \Omega_c^0 \, \pi^-$.

The systematic uncertainties on these mass measurements are similar to those
obtained for the $b$-mesons.
The mass scale uncertainty \cite{CDF_B_mass},
the choice of resolution model, and 
the uncertainty of the rest masses of the daughter
particles feed into the systematic uncertainties.
These effects are
combined in quadrature for total systematic uncertainties.  
It should be noted that momentum scale uncertainty will drop out
for measurements of the mass differences between the 
$\Xi_b^-$ and $\Xi_b^0$.  

Systematic uncertainties on the mean life measurements of the $B$
baryons are taken from the consistency observed in the $B$ meson
measurements, where we find complete consistency with nominal values
is determined to within $\pm6 \, \mu$m, or 1.3\%.  
Final results for the properties of the $B$ baryons are listed in 
Table \ref{table:final_prop}.

\begin{table}[hbt]
\begin{center}
\caption{$\Xi_c$ and $B$ Baryon Mass and Mean Life Results 
\protect \label{table:final_prop}}
\vspace{3mm}
\begin{tabular}{cccc}
\hline \hline
 Final State &  Mass (MeV/$c^2$) & Mean life ($ps$) \\
\hline
$\Lambda_b$                   & $5620.14\pm0.31(\textrm{stat})\pm0.40(\textrm{syst})$ &
 $1.565\pm0.035(\textrm{stat})\pm0.020(\textrm{syst})$ \\
$\Xi_b^- (J/\psi \, \Xi^-)$   & $5794.1\pm2.0(\textrm{stat})\pm0.40(\textrm{syst})$ &
 $1.36\pm0.15(\textrm{stat})\pm0.02(\textrm{syst})$ \\
$\Xi_b^- (\Xi^0_c \, \pi^-)$  & $5796.5\pm4.7(\textrm{stat})\pm0.95(\textrm{syst})$ & - \\
$\Xi_b^0$                     & $5791.6\pm5.0(\textrm{stat})\pm0.73(\textrm{syst})$ & - \\
$\Omega_b^- (J/\psi \, \Omega^-)$                  & $6051.4\pm4.2(\textrm{stat})\pm0.50(\textrm{syst})$ &
 $1.77^{+0.55}_{-0.41}(\textrm{stat})\pm0.02(\textrm{syst})$ \\
$\Omega_b^- (\Omega^0_c \, \pi^-)$   & $6040\pm8(\textrm{stat})\pm2(\textrm{syst})$ & -  \\
\hline
\hline
\end{tabular}
\end{center}
\end{table}

\section{Conclusions \label{sect:Conclusions}}
In conclusion, the full CDF data set has been analyzed 
to identify the largest possible
sample of ground state $B$ baryons.  The mass and mean life properties
of these particles have been measured, and the results compared to very 
precisely measured quantities for $B$ mesons obtained in similar final states.
The first evidence for the process 
$\Omega_b^- \rightarrow \Omega_c^0 \, \pi^-$ has been shown.
The final results of the mass and mean life measurements are listed in 
Table  \ref{table:final_prop}.  
The isospin splitting of the $\Xi_b^{-,0}$ states is found to be
$M(\Xi_b^-)-M(\Xi_b^0)$ =  
$2.5\pm5.4(\textrm{stat})\pm0.6(\textrm{syst})$ MeV/$c^2$.
These results supersede previous measurements 
which were obtained with a smaller data set \cite{CDF_Omega_b},\cite{CDF_Xib0}
and are consistent with recent results from LHCb \cite{LHCb_Xib}.

We thank the Fermilab staff and the technical staffs of the
participating institutions for their vital contributions. This work
was supported by the U.S. Department of Energy and National Science
Foundation; the Italian Istituto Nazionale di Fisica Nucleare; the
Ministry of Education, Culture, Sports, Science and Technology of
Japan; the Natural Sciences and Engineering Research Council of
Canada; the National Science Council of the Republic of China; the
Swiss National Science Foundation; the A.P. Sloan Foundation; the
Bundesministerium f\"ur Bildung und Forschung, Germany; the Korean
World Class University Program, the National Research Foundation of
Korea; the Science and Technology Facilities Council and the Royal
Society, UK; the Russian Foundation for Basic Research; the Ministerio
de Ciencia e Innovaci\'{o}n, and Programa Consolider-Ingenio 2010,
Spain; the Slovak R\&D Agency; the Academy of Finland; the Australian
Research Council (ARC); and the EU community Marie Curie Fellowship
contract 302103.

\end{document}